\documentstyle[12pt,epsf,amssymb]{article}

\def\be{\begin{eqnarray}}
\def\ee{\end{eqnarray}}
\def\bo{\begin{eqnarray*}}
\def\eo{\end{eqnarray*}}

\def\He#1{{}$^{#1}${\rm He}}
\def\eps{\varepsilon}
\def\lam{\lambda}

\def\k{{\bf k}}

\def\boldnabla{\mbox{\boldmath{$\nabla$}}}
\def\hpi{\hat\pi}
\def\hxi{\hat\xi}

\def\rr{{\bf r}}

\def\Grr#1#2{\langle\!\langle \rho_{#1}\,| \,\rho_{#2}\rangle\!\rangle}

\def\Grp#1#2{\langle\!\langle \rho_{#1}\,| \,\hpi_{#2}\rangle\!\rangle}

\def\grr{G_{\rho\rho}}
\def\gpr{G_{\pi\rho}}
\def\grp{G_{\rho\pi}}
\def\gpp{G_{\pi\pi}}
\def\gxr{G_{\xi\rho}}

\def\opA{\,\,\hat{\!\!\cal{{A}}\,}}

\begin{document}

\title{Helium-4 Energy and Specific Heat\\ in Superfluid and Normal Phase}
\author{A.~A.~Rovenchak}

\maketitle

\begin{abstract}
The calculation of the \He4 energy and specific heat is carried
out in a wide temperature range within the two-time temperature Green functions
approach. The approximation improving the random phase
approximation is developed providing the correct behaviour of
the calculated specific heat at the temperatures $0 < T/T_c < 0.65$
and $T/T_c> 1.32$ where $T_c$ stands for the phase transition
point.


PACS number(s): 05.70.Ce, 67.20.+k, 67.40.-w.
\end{abstract}

\section{Introduction}
The problem of the theoretical description for the many-boson
system thermodynamics in a wide temperature range is a very
curious one and still remains unsolved. Liquid \He4 is the best example for the
theretical studies because of its anomalous low-temperature
behaviour. It is well-known how to
obtain the thermodynamic functions in the limit $T\to0$
(this is given by the theory of Bogoliubov~\cite{Bogoliubov}). The renormalization
group theory also gives a fairly well results at the point of the
lambda-transition~\cite{Kleinert}. But still, quite a large temperature range
might be studied by means of numerical computations only.

Sears~\cite{Sears83} utilized neutron scattering results in order
to determine the internal energy of liquid helium-4 via some
simple thermodynamic relations. For these calculations, the
potential of Aziz {\it et al}~\cite{Aziz79} was used for representing the
He--He interatomic interaction.

Path Integral Monte Carlo (PIMC) technique which allows for the
study of the physical quntities temperature dependences was utilized
by Ceperley and Pollock~\cite{Ceperley86} for the helium-4 thermodynamic properties
calculation at the temperatures above 1~K.
Kinetic energy of liquid and solid helium-4 was calculated
in~\cite{Ceperley96} at high temperatures.
Such numerical simulations are numerous and we do not intend to analize them
here. The overview of Path Integral Monte Carlo (PIMC) techniques application
in the theory of condensed helium is adduced in the paper by Ceperley~\cite{Ceperley95}.

The derivation of the analytical relations for the thermodynamic
properties was made by Vakarchuk~\cite{Vakarchuk}. In the first of
the cited works, the expression for the many-boson two-particle density matrix
was written. The second work contains the analysis of the
non-linear density fluctuations contribution into the
thermodynamic properties of the Bose-system. The expressions found
there are expected to provide a good description in a wide
temperature range.

In this work, we will proceed from a very well studied random phase
approximation (RPA) which allows one to receive a low-temperature
asymptotics for the thermodynamic and structure functions of liquid
\He4. Then, making some quite natural assumptions, we will develop
what might be called `effective post-RPA' since the post-RPA
corrections will be taken into account in an effective manner. In
principle, such a technique will move us substantially ahead form the
low-temperature region. But unfortunately the way of the equations
solution applied here does not allows to grasp the temperatures
$|T/T_c-1|\lesssim1/3$ where $T_c$ stands for the
lambda-transition point, this seems to be rather a technical problem.

In this paper, we use the previously obtained potential~\cite{JPS00} for
the calculation of thermodynamic properties of liquid helium. A
two-time temperature Green function technique will be utilized for
the calculation of the energy and specific heat. The application
of this technique is well-known, it was used by Tserkovnikov~\cite{Tserkovnikov} for the
non-ideal Bose-gas in the random phase approximation.
Additionally, this method leads to a quite short way to the RPA (Bogoliubov's)
results. Unlike what is usualy done, the collective variables being
the density fluctuations Fourier
components are utilized for constructing the Green functions
instead of the creation--annihilation operators.

The organization of the paper is as follows. In Section~II, we introduce
the operators for the Green functions and write the Hamiltonian. In
Section~III, the equations of motion are written and solved. Next, the
function $\gamma_k(T)$ for the accounting of post-RPA terms is
introduced. The equations for it are adduced in Section~IV. Also,
the expression for the free energy is written and other thermodynamic
functions are derived from it in this Section. The discussion is
presented in Section~V.

\section{Collective variables and Hamiltonian}
We consider operators
\be     \label{rpxdef}
\rho_\k&=&{1\over\sqrt{N}} \sum_{j=1}^N e^{-i\k\rr_j},\\
\hpi_\k&=&{1\over\sqrt{N}}
   \sum_{j=1}^N e^{-i\k\rr_j}\left({-i\hbar^2\over m}\k\boldnabla_j\right),\\
\hxi_\k&=&{1\over\sqrt{N}}
   \sum_{j=1}^N e^{-i\k\rr_j}\left({-i\hbar^2\over m}\k\boldnabla_j\right)^2,
\ee
the operators $\rho_\k$ being collective variables, operators $\hpi_\k$
 are usually known as `currents'.
No special name will be given to $\hxi_\k$.

The Hamiltonian $\hat H$ of the system is written as
$\hat H= \hat{H}^{(0)} + \Delta\hat H$ where $\hat H^{(0)}$ corresponds
to RPA and $\Delta\hat H$ is the respective correction.
Hereafter we will consider only RPA term and therefore use $\hat H$
to designate $\hat H^{(0)}$:
\be
\hat H=\hat K +\hat\Phi
\ee
where $\hat K$ and $\hat\Phi$ being the kinetic and potential energy
in RPA respectively.

We have the following expression for the potential energy operator in
$\rho$-representation:
\be     \label{Phidef}
\hat\Phi={N(N-1)\over 2V}\nu_0
+\sum_{\k\neq0}{N\over 2V}\nu_{k}\left(\rho_{\k}\rho_{-\k}-1\right),
\ee
where $\nu_k$ means the Fourier transformation of the interatomic
potential in helium~\cite{JPS00},
while the kinetic energy is given in a usual way:
\be     \label{Kdef}
\hat K=-{\hbar^2\over 2m}\sum_{j=1}^N \boldnabla_j^2 .
\ee
It is interesting as well to consider the $\rho$-representation for
$\hat K$. Neglecting terms with two sums over the wave vector (i.~e.,
leaving the RPA term only) we obtain
\be     \label{Krhodef}
\hat K_\rho=\sum_{\k\neq0}\left\{
                   {1\over2}\rho_{-\k}\hpi_{\k}
                  -{1\over4\eps_{k}}\hpi_{-\k}\hpi_{\k}
                  \right\},
\ee
$\eps_k$ is the energy of a free particle:
\bo
\eps_k={\hbar^2 k^2 \over 2m}.
\eo

\section{Equations of motion for Green functions}
Since the total energy of the system $E$ is the average value of the
Hamiltonian we need to know the way of averaging our expressions.
In order to find the average values of such expressions as
$\rho_{\k_1}\rho_{\k_2}$,
$\hpi_{\k_1}\rho_{\k_2}$, etc. we will use two-time temperature Green
functions~\cite{Zubarev}:
\be
\langle\!\langle A(t)|B(t')\rangle\!\rangle=i\theta(t-t')\langle[A(t),B(t')]\rangle
\ee
with operators given in the Heisenberg representation, $\theta$ is
the Heaviside step function.
The following designations will be used:
\be
\grr(\k_1, \k_2)\equiv\Grr{\k_1}{\k_2},\quad
\grp(\k_1, \k_2)\equiv\Grp{\k_1}{\k_2}, \quad \ldots \, .
\ee
The averaging might be symbolically represented by an operator $\opA$:
\be
\langle AB \rangle = \opA G_{BA}.
\ee

We consider here only the case of $\k_2=-\k_1\equiv\k$ since such
expressions appear in our Hamiltonian~(\ref{Phidef}), (\ref{Krhodef}).
Therefore, the indexing of $\k_1$, $\k_2$ will be skipped for the sake of
simplicity.

\subsection{$\rho$-representation for the kinetic energy}

It is easy to show that one can obtain the following equations of motion
\be \label{System-rho}
\hbar\omega\,\grr(\omega)&=&-\eps_k\,\grr(\omega)+\gpr(\omega),\nonumber\\
\hbar\omega\, \gpr(\omega)&=&{\phantom-}\eps_k\,\gpr(\omega)
+\eps_k^2(\alpha_k^2-1)\,\grr(\omega)+{1\over\pi}\eps_k,
\ee
where
\be \label{alphadef}
\alpha_k=\sqrt{1+{2\varrho\nu_k\over \eps_k}},\qquad \varrho={N\over V}.
\ee
Using the solutions of the above system we receive the following
mean values
\be  \label{pairAverages}
\langle\rho_{-\k}\rho_{\k}\rangle&\equiv&\opA\grr(\k)\equiv S_k = {1\over\alpha_k}\coth{\eps_k\alpha_k\over2T},\nonumber\\
\langle\rho_{-\k}\hpi_{\k}\rangle &\equiv&\opA\gpr(\k)\equiv {\sl\widetilde\Pi}_k = {1\over2}\left({1\over\alpha_k}\coth{\eps_k\alpha_k\over2T}-1\right)
={1\over2}\left(S_k-1\right),\nonumber\\
\langle\hpi_{-\k}\hpi_{\k}\rangle&\equiv&\opA\gpp(\k)\equiv {\sl\Pi}_k = {1-\alpha_k^2\over4\alpha_k}\coth{\eps_k\alpha_k\over2T}
={1-\alpha_k^2\over4}\,S_k.
\ee
Here, the average $\langle\rho_{\k}\rho_{-\k}\rangle$ is nothing but the structure
factor of the system.

\subsection{Exact kinetic energy}

In this case, the equations of motion read:
\be
\hbar\omega\,\bar\grr(\omega)&=&-\eps_k\,\bar\grr(\omega)+\bar\gpr(\omega),\nonumber\\
\hbar\omega\,\bar\gpr(\omega)&=&-\eps_k\,\bar\gpr(\omega)
+\eps_k^2(\alpha_k^2-1)\,\bar\grr(\omega)+\bar\gxr(\omega)+{1\over\pi}\eps_k.
\ee

Unlike the previous case when the kinetic energy operator was
given in the $\rho$-representation, now it turned to be imposible
to solve this system with respect to $\bar\grr$ even in the RPA.
The problem lies in the appearence of the functions with the
higher orders of $(-i\hbar^2\boldnabla/m)$ in each equation following the
equation for $\bar\gxr$, etc., compare with (\ref{System-rho}).
Therefore, one should apply some approximation in order to obtain
the closed system of the equations of motion.

Let us consider the system of two equations for the functions
$\bar\grr$ and $\bar\gpr$ making a suggestion that
\be
\bar\gxr-2\eps_k\,\bar\gpr = 2\gamma_k(T)\eps_k^2\,\bar\grr,
\ee
with the factor of $\gamma_k(T)$ being of the energy dimension.
Its form will be choosed further.

Such a substitution returs us to system (\ref{System-rho}) but
the quantity $\alpha_k$ looks different now, we will refer it as
$\bar\alpha_k$:
\be
\bar\alpha_k &=&
     \sqrt{ 1+{2\varrho\nu_k \over \eps_k} + {2\gamma_k(T)\over \eps_k} }
\ee
One should keep in mind that a bar over a letter means `exect with respect to RPA'.

\subsection{General properties of $\gamma_k(T)$.}

Let us consider the cases of low and high temperatures.

In the low-temperature limit the results obtained in $\rho$-representation
are correct (the theory of Bogoliubov). Thus,
\be             \label{gammalow}
\gamma_k(T)\biggr|_{T\to0} = 0.
\ee
In the high-temperature limit, one has the expression for the structure
factor:
\be
S_k(T)\biggr|_{T\to\infty} = {1\over 1 + \varrho\nu_k/T }.
\ee
It leads to
\be                     \label{gammahigh}
2\gamma_k(T)\biggr|_{T\to\infty} = 2 T = {4\over3}K,
\ee
where $K$ is the kinetic energy per particle, in the high-temperature limit it equals
to the classical expression $K=3T/2$.

\section{Equation for $\gamma_k(T)$ and calculation of energy}
One can calculate the free energy of the system using the
expression~\cite{Vakarchuk}
\be \label{FreeEndef}
F=F_0 + \int_0^1 d\lam\, \Phi(\lam),
\ee
$F_0$ is the free energy of non-interacting system, $\lam$ is the
`interaction turn-on' parameter, $\Phi$ is the potential energy
given by
\be
{\Phi\over N} = {\varrho\nu_0\over 2}+{1\over 2N}\sum_{\k\neq0}
\varrho\nu_k \left({1\over\bar\alpha_k}\coth{\beta\over2}\eps_k\bar\alpha_k-1\right)
\ee

In order to integrate $\Phi(\lam)$ in (\ref{FreeEndef}) we will use the
integration over $\bar\alpha_k$ instead of $\lam$. This involves the
derivative $\partial \gamma_k / \partial\lam$. One cannot found it without knowing the explicit
dependence $\gamma_k(\lam)$. In this situation, the most simple
way is to use the method from~\cite{Bodies} and to write the so
called first difference:

\be \label{Suggestion1}
   {\partial \gamma_k \over \partial\lam}={\gamma_k-\gamma^0_k\over\Delta\lam}=\gamma_k-\gamma^0_k
\ee
since $\Delta\lam=1$. Here, $\gamma^0_k$ is the quantity
$\gamma_k$ for the ideal Bose-gas.

\subsection{Low temperatures, $T<T_\lam$}
A self-consistent relation for obtaining the quantity $\gamma_q(T)$ might be
sketched as follows:
\be  \label{textEq}
\fbox{$
\begin{array}{l}
{\rm Energy\ received}\\
{\rm from\ the\ averaged}\\
{\rm RPA{\textrm -}Hamiltonian}
\end{array}
$}
\qquad =\qquad
\fbox{$
\begin{array}{l}
{\rm Energy\ calculated}\\
{\rm from\ the\ free\ energy}\\
{\rm as}\ \ E=\displaystyle{\partial \;\beta F\over\partial\beta}
\end{array}
$}\ .
\ee
One should remember that every RPA-expression contains $\bar\alpha_k$ with $\gamma_k(T)$.

The left-hand side of the equation (\ref{textEq}) equals
\be    \label{EnRPA}
{E\over N}={\varrho\nu_0\over2}+{1\over N}\sum_{\k\neq0}
 \left\{\left({1\over2}{\sl\widetilde\Pi}_k-{1\over4\eps_k}{\sl\Pi}_k\right)+{1\over2}\varrho\nu_k(S_k-1)\right\},
\ee
where relation (\ref{pairAverages}) is taken into account.

One has for the free energy:
\be  \label{FreeEngamma}
{F\over N}&=&{F_0\over N}+{\varrho\nu_0\over2}
+{1\over 2N}\sum_{\k\neq0}\varrho\nu_k\left\{
{\eps_k(\bar\alpha_k-\bar\alpha^0_k)\over\varrho\nu_k+\gamma_k-\gamma^0_k}-1+{}\right.\nonumber\\
&&\qquad\left.{}+{2T\over\varrho\nu_k+\gamma_k-\gamma^0_k}
\ln{1-\exp\left(-\beta\eps_k\bar\alpha_k\right)\over1-\exp\left(-\beta\eps_k\bar\alpha^0_k\right)}
\right\}.
\ee
As a result, relation (\ref{textEq}) is rewritten as follows:
\be \label{En=En}
&&{\varrho\nu_0\over2}+{1\over 2N}\sum_{\k\neq0}\left[
\left(\eps_k\alpha_k - {\gamma_k\over\bar\alpha_k}\right)\coth{\eps_k\bar\alpha_k\over 2T}
-\eps_k -\varrho\nu_k\right]=
\nonumber\\
&&\qquad=K_0(T)+{\varrho\nu_0\over2}+{1\over 2N}\sum_{\k\neq0}
\varrho\nu_k\left\{
{2(\gamma^0_k{}'-\gamma_k')\over (\varrho\nu_k+\gamma_k-\gamma^0_k)^2}
   \ln{\sinh{\beta\over2}\eps_k\bar\alpha_k\over\sinh{\beta\over2}\eps_k\bar\alpha^0_k}
+{}\right. \nonumber\\
&&\qquad{}+{\eps_k\over \varrho\nu_k+\gamma_k-\gamma^0_k}
  \left[
  \left(\bar\alpha_k-\beta\bar\alpha_k'\right)\coth{\beta\over2}\eps_k\bar\alpha_k
  -
  \left(\bar\alpha^0_k-\beta\bar\alpha^0_k{}'\right)\coth{\beta\over2}\eps_k\bar\alpha^0_k
  -1
  \right]
\left.\vphantom{\ln{\sinh{\beta\over2}\eps_k\bar\alpha_k\over\sinh{\beta\over2}\eps_k\bar\alpha^0_k}}\right\},
\ee
where primes mean derivation over $\beta$ and
\be
\bar\alpha^0_k=\sqrt{1+{2\gamma^0_k(T)\over \eps_k}}.
\ee
For the $\bar\alpha_k$, $\bar\alpha^0_k$ derivatives we obtain:
\be  \label{alphaprime}
\bar\alpha_k'={\gamma_k'\over\eps_k\bar\alpha_k}, \qquad
\bar\alpha^0_k{}'={\gamma^0_k{}'\over\eps_k\bar\alpha^0_k}.
\ee

The quantity $\gamma^0_k$ still remains unknown. We will try to
find it utilizing the thermodynamic relations for the ideal Bose-gas.

Let us consider an arbitrary (no matter, interacting or non-interacting) system first.
Its structure factor in the long-wave limit reads:
\be
S_k(T)\bigg|_{k\to0}=\lim_{k\to0}{1\over\bar\alpha_k}\coth{\beta\over2}\eps_k\bar\alpha_k=
{T\over \eps_0+\varrho\nu_0+\gamma_0}.
\ee
The expression $\eps_0$ obviously equals to zero but we reserve it
in the sence of the limit $k\to0$, this will be necessary for the ideal gas case.
Therefore,
\be  \label{gamma0}
\gamma_0={T\over S_0(T)}-\varrho\nu_0-\eps_0={1\over\varrho\varkappa_T}-\varrho\nu_0-\eps_0,
\ee
where the connection between structure factor at $k=0$ and the
isothermic compressibility $\varkappa_T$ is taken into account. For the ideal gas $\nu_k=0$,
thus at low temperatures one receives:
\be
\gamma^0_0=\left(\partial P_0\over\partial\varrho\right)_T =0 \quad {\rm for}\quad T<T_c,
\ee
where $P_0$ is the ideal Bose-gas pressure and $T_c$ is the
Bose-condensation temperature. As a result, the ideal Bose-gas
structure factor $S^0_k$ reads:
\be
S^0_k(T)\biggr|_{k\to0}={2\over \eps_k}\propto {1\over k^2},
\ee
as it should be~\cite{1/q^2}.

Now, we accept for the simplicity that the dependences on the wave
vector and the temperature in quantities $\gamma_k$ and $\gamma^0_k$ separate:
\be \label{Suggestion2}
\gamma_k(T)=\gamma(T) f_k,\quad \gamma^0_k(T)=\gamma^0(T) f^0_k,\qquad
f_0=f^0_0=1.
\ee
Te above statement means the separation of the `primary'
temperature dependence in the comparison with the `secondary'
dependence on the wave vector. The latter will be integrated up
when calculating the thermodynamic properties. The condition $f_0=f^0_0=1$ is nothing
but the specific normalization condition.

Now, from $\gamma^0_0=0$ one obtains $\gamma^0=0$ and, taking (\ref{alphaprime}) into account,
also $\gamma^0_k{'}=0$, $\bar\alpha^0_k{'}=0$. These equalities are
valid for $T<T_c$.

Let us consider the expansion of (\ref{En=En}) in terms of
$\gamma$ in order to obtain the equation for $\gamma_k(T)$.
The first-order terms read:
\be  \label{uptogamma2}
&&T\dot\gamma(T) \int_0^\infty k^2 f_k\left\{
{2T\over\varrho\nu_k}\ln{\sinh{\eps_k\alpha_k\over2T}\over\sinh{\eps_k\over2T}}
-{1\over\varrho\nu_k}\coth{\eps_k \alpha_k\over 2T}
\right\}+{}
\\&&{}+
\gamma(T)\int_0^\infty k^2 f_k\left\{
{1\over\alpha_k}\coth{\eps_k \alpha_k\over 2T}
+{\eps_k\over\varrho\nu_k}\left[\coth{\eps_k\over 2T}- \alpha_k\coth{\eps_k \alpha_k\over 2T}\right]
\right\} =0,\nonumber
\ee
where the dot denotes the derivative with respect to the temperature and
$\alpha_q$ is defined by the Bogoliubov's expression (\ref{alphadef})

Similar expansion for energy (\ref{EnRPA}) leads to the following equation:
\be
E&=&{1\over4\pi^2\varrho}\int_0^\infty dk\,k^2 \eps_k \left[
   \alpha_k\coth{\eps_k\alpha_k\over 2T}-{\alpha_k^2+1\over2}\right]
    -\gamma{1\over4\pi^2\varrho}\int_0^\infty dk\,k^2
    f_k {1\over 2 T}\eps_k{1\over\sinh^2{\eps_k\alpha_k\over 2T}} \\
 &+&\gamma^2{1\over4\pi^2\varrho}\int_0^\infty dk\,k^2\,
   f_k^2
   \Biggl\{
   {1\over 4T{\alpha_k}}
    \left[{1\over \alpha_k}+{\eps_k\coth{\eps_k\alpha_k\over 2T}\over T}
    \right]
      {1\over\sinh^2{\eps_k\alpha_k\over 2T}}
  +{1\over 2T\eps_k{\alpha_k}^3}\coth{\eps_k\alpha_k\over 2T}
  \Biggr\} \nonumber
\ee

The last item in the second integral causes some difficulties since it does not
converge at $k\to\infty$. One can ensure the convergence by means of the function $f_k$
(it is arbitrary by now).
The next problem with this item is the incorrect behaviour of the heat
capacity at low temperatures. As it will be shown further,
$\gamma(T)$ is proportional to $T$ when $T\to0$.
It leads to the linear temperature dependence of heat capacity
while one should have $\propto T^3$. We can avoid this problem by choosing
$f_k$ in the special form. We will demand the following:
\be                 \label{fq2eq0}
\int^\infty_0 dk\, k^2\, f_k^2 \,
  {1\over\eps_k\alpha_k^3}=0.
\ee
The similar problem is with $\gamma^3$ giving quadratic dependence
with respect ot $T$. Thus, the second equation:
\be             \label{fq3eq0}
\int^\infty_0 dk\, k^2\, f_k^3 \,
  {1\over\eps_k^2\alpha_k^5}=0.
\ee

Only a complex $f_k$ can satisfy both these equation.
For the sake of simplicity in calculations we have choosed $f_k$ in the
following form:
\be
f_k=\left\{
\begin{array}{cl}
  1,           & 0  \le k<k_1,\\
 if_{\rm II},  & k_1\le k<k_2,\\
- f_{\rm III}, & k_2\le k<k_3,\\
-if_{\rm IV},  & k_3\le k<k_4,\\
0, & {\rm otherwise}
\end{array}
\right.
\ee
with equally spaced intervals, $k_n=n\times2$~\AA$^{-1}$.
Three quantities $f_{\rm II}, f_{\rm III}, f_{\rm IV}$ are necessary
to satisfy both real and complex terms in the equations
(\ref{fq2eq0}--\ref{fq3eq0}).

The equation for $\gamma$ might be rewritten as follows:
\be
a_1(T)\,T\dot\gamma+a_0(T)\gamma+O(\gamma^2)=0
\ee

\begin{figure}[h]
\epsfxsize=75mm
\centerline{\epsfbox{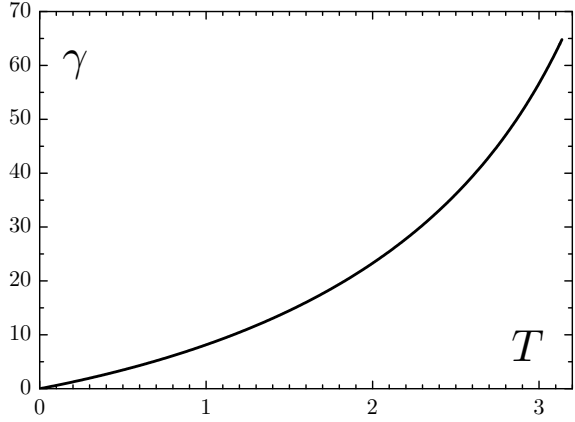}}
\caption{The dependence of $\gamma$ on the temperature.
}   \label{gamma6}
\end{figure}

Neglecting higher-order terms one obtains:
\be
\gamma(T)=CT\exp\left\{-\int^T {d\tau\over \tau}
                     \left[{a_0(\tau)\over a_1(\tau)}+1\right]
                \right\},
\ee
where the constant $C$ is unknown yet. One can define it studying
the behaviour of the isothermic compressibility near the absolute
zero (\ref{gamma0}). We received the following value by averaging
the results of the data~\cite{NIST} interpolation at the lowest
temperatures over four, five and six points:
\be
C\equiv\dot\gamma(0)\simeq6.6 \,.
\ee
The $\gamma(T)$ curve is adduced at Fig.~\ref{gamma6}.

\bigskip\bigskip

One should notice that teh correct results must be expected only
for the temperatures at which it is possible to neglect the higher-order terms
in the equation for $\gamma_k$ (\ref{uptogamma2}).
In other words, if $2\gamma_k/\eps_k>1$ then the results are not
correct. Taking into account that the typical wave vector absolute
value for \He4 is 2~\AA\ (it corresponds to the first maximum on the structure factor curve)
we obtain the upper limit for the applicability of our results as $T/T_c\simeq 2/3$.

\subsection{High temperatures, $T>T_\lam$}

In this case one can use the techniques similar to the described
in the previous sections for low temperatures. Unfortunately, this
seems to be a rather complicated problem. Firstly, the function
$\gamma^0_k(T)$ is not equal to zero already. Secondly,
the quantity $\gamma_k(T)$ is not small in this temperature region
and it is not possible to use the expansion into series, we also did not manage to solve
equation (\ref{En=En}) (the initial condition for the differential equation is unknown).
Therefore, for the simplification of the calculations
we decided to substitute $\gamma_k(T)$ with $2K/3$, see (\ref{gammahigh}),
where $K$ is the kinetic energy per particle.

We have rewritten the expression for the free energy (\ref{FreeEngamma}) in order to obtain
the equation for $\Delta K \equiv K-K_0$:
\be
\Delta K &=& {1\over 4\pi^2} \int_0^\infty dk\,k^2\  \nu_k
  \Biggl\{
    {1\over Z_k}\eps_k
      \left[
        \alpha_k \coth {\beta\over 2}\eps_k\alpha_k
       -\alpha_k^0 \coth {\beta\over 2}\eps_k\alpha_k^0
      \right]
      -{1\over \alpha_k}\coth{\beta\over2}\eps_k\alpha_k \nonumber\\
  &&\qquad
     +{2\over3}\,{\partial\Delta K\over\partial T}
      \Biggl[
        {T\over Z_k^2}\eps_k(\alpha_k-\alpha_k^0)
       -{T\over Z_k}{1\over\alpha_k}\coth {\beta\over2}\eps_k\alpha_k\nonumber\\
  &&\qquad\qquad
      +{2T^2\over Z_k^2}\Biggl( \ln\left(1-e^{-\beta\eps_k\alpha_k}\right)
                               -\ln\left(1-e^{-\beta\eps_k\alpha_k^0}\right)
                         \Biggr)
      \Biggr] \nonumber\\
  &&\qquad
     +{2\over3}\dot K_0{T\over Z_k}
      \Biggl[
        {1\over\alpha_k^0}\coth{\beta\over2}\eps_k\alpha_k^0
       -{1\over\alpha_k}\coth{\beta\over2}\eps_k\alpha_k
      \Biggr]
   \Biggr\}, \label{DeltaKeq}\\
Z_k &=& \varrho\nu_+{2\over3}\Delta K, \nonumber\\
\alpha_k &=& \sqrt{ 1+{2\varrho\nu_k \over \eps_k}
                   +{4\over3}{\Delta K + K_0 \over \eps_k}
                  }, 
                  \nonumber\\
\alpha_k^0&=&\sqrt{ 1 + {4\over3}{K_0 \over \eps_k},
                  } 
                  \nonumber
\ee
where the dot means derivation with respect to $T$: $\dot K_0=\partial K/\partial T$.

\begin{figure}[h]
\epsfxsize=75mm
\centerline{\epsfbox{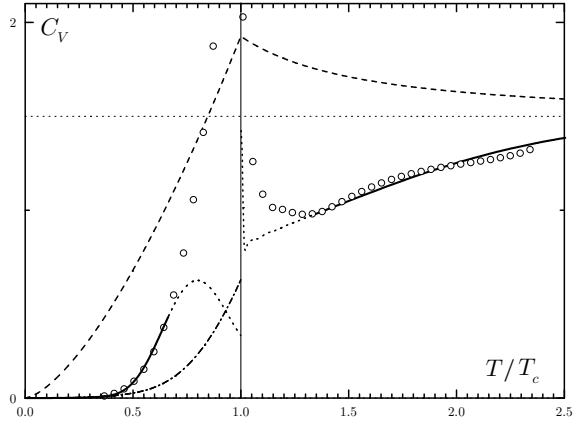}}
\bigskip
\caption{ \label{CVresults}
Energy of liquid helium-4. Solid line --- calculated energy,
dotted line shows the regions where the approximations used
cannot be accepted;
dashed line --- energy of ideal Bose-gas;
dashed-dotted line --- Bogoliubov's approximation.
}
\end{figure}

Having calculated the kinetic energy form Eq.~(\ref{DeltaKeq}) we
then proceed to the calculation of the total energy. After the
numerical derivation of the latter we receive the specific heat of
the liquid helium. This result is adduced in Fig.~\ref{CVresults}
together with the data for the low temperatures from the previous
subsection.

\bigskip

\section{Discussion}

As one can see from Fig.~\ref{CVresults} we obtained a fairly good
description in the region $0\leq T/T_c<0.65$, as it was expected.
In the normal phase ($T>T_c$) a similar situation is received for
the temperatures $T/T_c>1.32$.
It is necessary to say that the phase transition temperature
$T_c$ in our case coincides with the respective temperature of the
ideal Bose-gas. One should use the effective mass $m^*$ of the helium
atom instead of that of free atom $m=4.0026$~a.~m.~u.
As will be shown elsewhere~\cite{MyEffMass}, it is possible to
receive the effective mass in a sel-consistent manner within the Green function formalism.
The ground-state ($T=0$~K) value equals to $m^*\simeq1.58\,m$.
We assume that the temperature dependence of the effective mass is
significant in the temperature region under consideration:
$m^*(T)\simeq {\rm const}=m^*(0)=1.58\,m$.
Therefore, one receives the
transition temperature of 1.99~K {\it vs} experimental 2.17~K~\cite{MyEffMass}.

\end{document}